 \documentclass[preprint,aps]{revtex4}

\usepackage{graphicx}
\usepackage{bm}

\begin{document}

\title{Evidence for Bosonic Mode Coupling in Electron Dynamics of LiFeAs Superconductor}

\author{Cong Li$^{1,2}$,  Guangyang Dai$^{1,2}$, Yongqing Cai$^{1,2}$, Yang Wang$^{1,2}$, Xiancheng Wang$^{1,2}$, Qiang Gao$^{1,2}$, Guodong Liu$^{1}$, Yuan Huang$^{1}$, Qingyan Wang$^{1}$, Fengfeng Zhang$^{3}$, Shenjin Zhang$^{3}$, Feng Yang$^{3}$, Zhimin Wang$^{3}$, Qinjun Peng$^{3}$, Zuyan Xu$^{3}$, Changqing Jin$^{1,2,4}$, Lin Zhao$^{1,*}$ and X. J. Zhou$^{1,2,4,5,*}$
}

\affiliation{
\\$^{1}$Beijing National Laboratory for Condensed Matter Physics, Institute of Physics, Chinese Academy of Sciences, Beijing 100190, China
\\$^{2}$University of Chinese Academy of Sciences, Beijing 100049, China
\\$^{3}$Technical Institute of Physics and Chemistry, Chinese Academy of Sciences, Beijing 100190, China
\\$^{4}$Songshan Lake Materials Laboratory, Dongguan 523808, China
\\$^{5}$Beijing Academy of Quantum Information Sciences, Beijing 100193, China
\\$^{*}$Corresponding authors: lzhao@iphy.ac.cn and XJZhou@iphy.ac.cn
}


\pacs{}

\begin{abstract}
Super-high resolution laser-based angle-resolved photoemission  measurements are carried out on LiFeAs superconductor to investigate its electron dynamics. Three energy scales at $\sim$20 meV, $\sim$34 meV and $\sim$55 meV are revealed for the first time in the electron self-energy both in the superconducting state and normal state. The $\sim$20 meV and $\sim$34 meV scales can be attributed to the coupling of electrons with sharp bosonic modes which are most likely phonons. These observations provide definitive evidence on the existence of mode coupling in iron-based superconductors.
\end{abstract}

\maketitle


The physical properties and superconductivity of materials are dictated by electron dynamics, in particular, how electrons interact with other particles or excitations, such as electron-electron interaction, electron-phonon coupling and electron-impurity scattering. In conventional superconductors, the electron-phonon interaction plays a dominant role in giving rise to electron pairing and superconductivity\cite{BCSTheory,WLMcMillan}.  In high temperature cuprate superconductors, the study of such many-body effects is crucial in understanding the anomalous normal state properties and  the mechanism of high temperature superconductivity.  With the advancement of high resolution angle-resolved photoemission spectroscopy (ARPES), it has become a powerful tool to directly probe the electron dynamics and many-body effects in materials\cite{ADamascelli_RMP,XJZhou_Springer,XJZhou_RPoP}. In high temperature cuprate superconductors,  ARPES has revealed an energy scale of $\sim$70 meV along the nodal direction\cite{PVBogdanov,PJohnson,AKaminski,Lanzarakink,ZhouNaturekink,KordyukKink,WTZhangKink},  another energy scale of $\sim$40 meV near the anti-nodal region\cite{GromkoANKink,KimANKink,CukANKink,JFHe_PRL}, the existence of high energy scale at 200$\sim$400 meV\cite{Ronning,Graf,Xie,Valla,Meevasana,WTZhangHEKink}, and extraction of normal and pairing Eliashberg functions\cite{XJZhouMul,LZhao_PRB,JBok_SA}.

Although the role of electron-phonon coupling\cite{LBoeri}, spin fluctuation\cite{KKuroki,IIMazin,AVChubokov} and orbital fluctuation\cite{HKontani,TSaito} in generating superconductivity in iron-based superconductors has been discussed\cite{LBoeri},  there have been few ARPES studies of many body effects reported in iron-based superconductors since their discovery in 2008\cite{AAKordyuk_PRB,HMiao_PRB}. Since many-body effects show up as subtle changes in the electron self-energy, their detection asks for super-high resolution and high statistics ARPES measurements on an isolated band with a relatively large bandwidth\cite{ADamascelli_RMP,XJZhou_Springer}.  Different from cuprate superconductors where mainly one Cu d$_{x^2-y^2}$ orbital is involved in low energy excitations,  all five Fe 3d orbitals participate in producing multiple bands in iron-based superconductors\cite{DJSingh,LZhaoCPL,Paglione_NP}. The coexisting two or three hole-like bands near the Brillouin zone center and two crossing electron-like bands near the zone corner hinder the investigation of many-body effects in iron-based superconductors.  In addition, the narrow bandwidth in most iron-based superconductors\cite{DLFeng2014ZRYe_PRX,XJZhou2016DFLiu_CPL} adds more difficulty in analyzing the electron dynamics.  Among all the discovered iron based superconductors, LiFeAs is by far the most promising candidate for studying the electron dynamics\cite{CQJin2008XCWang_SSC,MGuloy2008JHTapp_PRB}.   First, LiFeAs is a stoichiometric superconductor that is free from strong impurity scattering so sharp quasiparticle peaks can be observed in ARPES measurements\cite{BBuchner2010SVBorisenko_PRL}. Second, the two hole-like bands around the zone center are well separated with a relatively large distance, making it possible to isolate each of them\cite{BBuchner2010SVBorisenko_PRL,HMiao_PRB}. However, while some signatures of electron-boson coupling were suggested in the previous ARPES study of one-hole band in LiFeAs\cite{AAKordyuk_PRB}, the experimental precision is not sufficient to make definitive conclusions, as we will show in the present work.

In this paper, we carried out laser-based angle-resolved photoemission measurements with super-high instrumental resolution on the electron dynamics of LiFeAs.  Taking advantage of photoemission matrix element effect by using different light polarizations, we are able to isolate each of the two hole-like bands around the zone center. This makes it possible, for the first time, to identify definitively the mode coupling on both hole-like bands in LiFeAs.

High resolution angle-resolved photoemission measurements were carried out on our lab system equipped with a Scienta DA30L electron energy analyzer\cite{XJZhou2008GDLiu_RSI}. We used helium discharge lamp that provides a photon energy of 21.218 eV (helium I) to map out the overall Fermi surface of LiFeAs (Fig. 1a) and a vacuum ultraviolet (VUV) laser with the photon energy of 6.994 eV to zoom in on the Fermi surface and band structure around the zone center.  The laser polarization can be tuned with its electric field vector, E, along different directions, as shown in Fig. 1(b-d).  For the laser-ARPES measurements, the bandwidth of the laser is $\sim$0.26 meV, and the energy resolution of the electron energy analyzer is set at 1.5 meV, giving rise to an overall energy resolution of 1.52 meV.  The angular resolution is $\sim$0.3$^{\circ}$, corresponding to a momentum resolution of 0.0043 $\AA^{-1}$ for the photon energy of 6.994 eV. The Fermi level is referenced by measuring on a clean polycrystalline gold that is electrically connected to the sample. High-quality single crystals of LiFeAs with a superconducting transition temperature, T$_{c}$, of  $\sim$18 K were grown by the self-flux method.  The samples were cleaved {\it in situ} in vacuum with a base pressure better than $5\times10^{-11}$ Torr.

Figure 1a shows the overall Fermi surface mapping of LiFeAs measured with a photon energy of 21.218 eV (HeI $\alpha$).  Around the zone center, two hole-like Fermi surface sheets are observed, labeled as $\beta$ (blue circle) and $\gamma$ (pink circle) while around the zone corner, M point, two crossing electron-like Fermi surface sheets, labeled as $\delta$ (black circles) are observed. By taking high resolution laser-based ARPES measurements, we can zoom into the zone center region, as shown in Fig. 1(b-d) measured under three different polarization geometries. It is clear that the Fermi surface mappings of LiFeAs exhibit different spectral distribution under different polarization geometries. Under a given polarization geometry, there is also a dramatic spectral weight variation along the measured Fermi surface sheet.  These strong photoemission matrix element effects are consistent with the previous ARPES measurements\cite{BBuchner2010SVBorisenko_PRL, AAKordyuk_PRB,HMiao_PRB} which indicate that the $\beta$ Fermi surface is composed of $d_{xz}/d_{yz}$ orbitals while the $\gamma$ Fermi surface is composed of $d_{xy}$ orbital. In the laser ARPES measurements,  a small Fermi surface, labeled as $\alpha$ (dashed green circle in Fig. 1b), can be clearly observed around the zone center in Fig. 1b and 1e which is attributed to a topological surface state in LiFeAs\cite{SShin2019PZhang_NP}. However, there is no sign of the $\alpha$ Fermi surface in Fig. 1c. We note that, while the major electric field vector lies in the sample plane in Fig. 1(b-d), there is a small out-of-plane component of the electric field vector for Fig. 1b and Fig. 1d, but zero component for Fig. 1c.  Combining the matrix element effect analysis, we can determine that the $\alpha$ Fermi surface is dominated by the $p_{z}$ orbital.

Figure 1(e-g) shows the band structure of LiFeAs measured along the $\Gamma$-$X$ direction under the three polarization geometries corresponding to Fig. 1(b-d), respectively.  Under the polarization geometry in Fig. 1b, all the $\alpha$,  $\beta$ and $\gamma$ bands can be clearly observed simultaneously, as shown in Fig. 1e.  But under the polarization geometry in Fig. 1c, the $\alpha$ band disappears completely, and the $\gamma$ band is strongly suppressed, leaving the strong $\beta$ band that is well isolated (Fig. 1f). When the polarization geometry changes to the one in Fig. 1d, the $\beta$ band becomes nearly invisible while the $\alpha$ band is strongly suppressed  (Fig. 1g).  This leaves the $\gamma$ band strong and well isolated from the other two bands.  With the strong polarization dependence, the results in Fig. 1f and Fig. 1g provide an ideal platform to investigate the electron dynamics and many-body effects associated with the $\beta$ band and $\gamma$ band.


Selecting the polarization geometry of Fig. 1d,  we took high resolution laser-ARPES measurement on the $\gamma$ band of LiFeAs measured along $\Gamma$ - $X$ direction at 20 K, as shown in Fig. 2a. The measured band  is well isolated and extends to high binding energy.  Representative momentum distribution curves (MDCs) and photoemission spectra (energy distribution curves, EDCs) are shown in Fig. 2d and Fig. 2e, respectively. Sharp EDC peaks can be observed at the Fermi momentum (Fig. 2e). The MDCs at different binding energies show up as well-defined peaks (Fig. 2d) which can be fitted by a Lorenztian. From the fitted MDC peak position at different binding energies, we obtain the quantitative dispersion relation as shown in Fig. 2f and corresponding MDC width shown in Fig. 2h that is related to the imaginary part of electron self-energy.  Taking a linear line connecting the points at the Fermi level and at 100 meV (blue dashed line) and 50 meV (red dashed line), we can get the empirical real part of the electron self-energy, $Re\Sigma$, by subtracting the measured dispersion with the bare band, as shown in Fig. 2g (blue line and red line). Three features can be identified from the effective real part of electron self-energy, $Re\Sigma$, located at $\sim$20 meV, $\sim$34 meV and $\sim$55 meV, as marked by black arrows in Fig. 2g. These indicate that there are three energy scales involved in the electron dynamics of the $\gamma$ band.

It is known that, when there is an electron coupling with a sharp bosonic mode, it will produce a kink in dispersion, and a spectral dip in the photoemission spectra (EDCs)\cite{ADamascelli_RMP,XJZhou_Springer}.  In order to check on possible electron-sharp mode coupling, we take second derivative with respect to energy on the original data in Fig. 2a to get the image in Fig. 2b.  The second derivative processing helps in enhancing weak features in the original photoemission spectra. As seen in Fig. 2b, in addition to the main band where a kink can be more clearly seen as marked by the red arrow,  there appear two features on the right side of the main band, and one feature on the left side, as marked by black arrows. These are typical features that can be attributed to the electron coupling with two sharp bosonic modes.  In order to understand these features, we carried out a simulation of the single-particle spectral function by considering electron coupling with two sharp bosonic modes at 20 meV and 34 meV, and then take second derivative of the simulated data to get the image in Fig. 2c\cite{SEngelsberg,JFHe_PRL}. Such a simple simulation in Fig. 2c nicely captures the main features observed in Fig. 2b although there is some spectral weight difference. Therefore, the two energy scales, 20 meV and 34 meV observed in the effective electron self-energy (Fig. 2g) and the EDC second derivative image (Fig. 2b), can be attributed to electron coupling with two sharp bosonic modes for the $\gamma$ band in LiFeAs.

Our polarization-dependent laser-ARPES measurements also make it possible to fully isolate the $\beta$ band (Fig. 1f) that was not possible before. We took laser-ARPES measurements on the $\beta$ band with high resolution and high statistics, as shown in Fig. 3a. Similar to analyzing the $\gamma$ band in Fig. 2,  Fig. 3c and Fig. 3d show the corresponding MDCs and EDCs, respectively, from Fig. 3a. Sharp EDC peak can also be observed at the Fermi  momentum for the $\beta$ band (Fig. 3a). The MDCs at different binding energies can be well fitted by a Lorentzian, and the fitted MDC position gives the dispersion in Fig. 3e and the MDC width in Fig. 3g.  By subtracting an empirical bare band (blue line and red line) from the measured dispersion in Fig. 3e, we obtained the effective real part of electron self-energy shown in Fig. 3f (blue line and red line). It also exhibits three features at $\sim$20 meV, $\sim$34 meV and $\sim$55 meV (marked by prink, green and blue strips in Fig. 3f).  From Fig. 3b that is the EDC second derivative image of Fig. 3a, a kink in the main band is observed near 34 meV, as marked by the red arrow. Also the feature corresponding to the $\sim$34 meV mode coupling is clear as marked by the black arrow. But the feature corresponding to 20 meV mode coupling is not clear in Fig. 3b possibly because it is too weak that is below the noise level of the data.  Overall, similar to $\gamma$ band,  there are also three energy scales observed at similar energy positions for the $\beta$ band in LiFeAs.

Temperature-dependent measurements can provide further information on the origin of the observed three energy scales. Fig. 4a and 4d show the effective real part of electron self-energy measured for the $\gamma$ band and $\beta$ band, respectively, at different temperatures above and below the superconducting transition temperature of 18 K.  The corresponding MDC widths for the two bands are shown in Fig. 4b and 4e, and the EDCs at the Fermi momentum are shown in Fig. 4c and 4f, respectively.   As the temperature decreases, the three features in the effective real part of electron self-energy become more pronounced. For these two bands, the three energy scales are present both in the normal state and in the superconducting state (Fig. 4a and 4d). These indicate that they are not generated from superconducting transition. In principle, when there is a sharp mode coupling with a frequency of $\omega_0$ in the normal state, it is expected to shift to a higher energy, $\omega_0$+$\Delta$, in the superconducting state with $\Delta$ being the superconducting gap\cite{AWSandvik}. Since the related superconducting gap is small, the energy position change with superconducting transition is not obvious within our experimental uncertainty.

As the temperature decreases, the MDC width exhibits a decrease mainly near the Fermi level and the low binding energy region, as seen in Fig. 4b and 4e for both the respective  $\gamma$ band and $\beta$ band. This agrees with the EDC sharpening with decreasing temperature shown in Fig. 4c and 4f.  When the features get sharper at low temperature, we can see directly a dip in EDC corresponding to the $\sim$34 meV energy scale, as shown in Fig. 4c.  We note that, while the three energy scales observed for the $\gamma$ band and $\beta$ band are similar in energy positions, there are obvious differences between their electron dynamics. First, the EDC width for the $\beta$ band (Fig. 4f)  is much larger than that of the $\gamma$ band (Fig. 4c); at 12 K, the EDC width of the $\beta$ band is nearly twice that of the $\gamma$ band (insets in Fig. 4f and 4c).  This indicates that the $\beta$ band experiences much stronger scattering than that of the $\gamma$ band. Second,  the $\gamma$ band is more sensitive to temperature change than the $\beta$ band: the EDC for the $\gamma$ band exhibits an obvious sharpening with decreasing temperature, as seen in the insets of Fig. 4b and Fig. 4c. But the $\beta$ band shows much less change in EDCs (Fig. 4f) and MDC width (inset of Fig. 4e) within similar temperature change.

Our combined results of the electron self-energy (Fig. 2g and Fig. 3f), the comparison between the measured and simulated bands (Fig. 2b and 2c), and the observation of dip structure in EDCs (Fig. 4c)  provide clear evidence on the three separate energy scales at $\sim$20 meV, $\sim$34 meV and $\sim$55 meV in $\gamma$ and $\beta$ bands. The energy positions are different from those reported before\cite{AAKordyuk_PRB} and our clear identification of these energy scales is due to improved instrumental resolution and data statistics from our laser-ARPES measurements.  Now it comes to the discussion on the origin of these three energy scales.  From the comparison between the measured and simulated results (Fig. 2b and 2c),  the energy scales at  $\sim$20 meV and $\sim$34 meV can be attributed to the sharp bosonic modes that couple with electrons. The first obvious candidate of the bosonic modes is phonons. Six phonon modes are observed in Raman scattering measurements of LiFeAs\cite{MLTacon2012YJUm_PRB}. The $A_{1g}$ (As) mode with an energy of 20 meV and the $E_{g}$ (Fe) mode with an energy of 34 meV are consistent with the two energy scales we observed. We note that the highest phonon energy observed in LiFeAs is $\sim$42 meV\cite{MLTacon2012YJUm_PRB}. Therefore, the energy scale of $\sim$55 meV cannot be attributed to electron coupling with single phonon mode. The second possibility is the electron coupling to collective magnetic excitations.  Because there is no static magnetic ordering or long-range orbital ordering observed in LiFeAs\cite{BBuchner2010SVBorisenko_PRL,MBraden2012NQureshi_PRL,DReznik2016AMerritt_JSNM},  the energy scale of $\sim$55 meV cannot be induced by electron coupling with magnons. The third possibility is the electron coupling with multiple phonons; in this case, the feature at $\sim$55 meV may be understood\cite{MLTacon2012YJUm_PRB}.  It is also possible that the feature at $\sim$55 meV may be induced by electron-electron interaction or electron coupling with some high energy excitations in LiFeAs. Further experimental and theoretical efforts are needed to pin down on the origin of this $\sim$55 meV energy scale.

We can further estimate the coupling strength of the observed energy scales.  Since the $\sim$20 meV and $\sim$34 meV scales are attributed to electron-phonon coupling, and the energy difference of the two modes is small, we choose to estimate their combined coupling strength. To do this, we took the Fermi velocity near the Fermi level v$_F$ (0$\sim$15 meV) and the velocity at a high binding energy v$_H$ (40$\sim$50 meV) from the measured dispersions for the $\gamma$ band (Fig. 2f) and $\beta$ band (Fig. 3e). The combined coupling strength, $\lambda_{ph}$, of the  $\sim$20 meV and $\sim$34 meV scales can be estimated from $\lambda_{ph}$=v$_H$/v$_F$-1. The obtained values of $\lambda_{ph}$ are similar for both the $\gamma$ and the $\beta$ bands that are $\sim$0.5. We did similar analysis for the dispersions measured at different temperatures and found that the coupling strength $\lambda_{ph}$ shows little change with temperature in the range of 12$\sim$25 K.


In summary, by carrying out high resolution laser-based ARPES on LiFeAs, we have observed three energy scales at $\sim$20 meV, $\sim$34 meV and $\sim$55 meV for the first time for both the $\beta$ and $\gamma$ bands in both the superconducting state and the normal state.  The energy scales at $\sim$20 meV and $\sim$34 meV are due to electron coupling with sharp bosonic modes which are most likely phonons in LiFeAs. The origin of the energy scale at $\sim$55 meV asks for further investigations. As in conventional superconductors, the identification of energy scales in the tunneling spectrum played a key role in identifying the pairing glue for superconductivity\cite{WLMcMillan},  the identification of clear energy scales will provide important information to study the role of the electron-boson coupling and identify the pairing glues in iron based superconductors.

\vspace{3mm}

\noindent {\bf Acknowledgment}\\
We thank financial support from the National Key Research and Development Program of China (Grant No. 2016YFA0300300, 2016YFA0300600,  2017YFA0302900, 2018YFA0704200, 2018YFA0305600 and 2019YFA0308000), the National Natural Science Foundation of China (Grant No. 11888101, 11922414 and 11874405), the Strategic Priority Research Program (B) of the Chinese Academy of Sciences (Grant No. XDB25000000 and XDB33010300), the Youth Innovation Promotion Association of CAS (Grant No. 2017013), and the Research Program of Beijing Academy of Quantum Information Sciences (Grant No. Y18G06).

\vspace{3mm}

\noindent {\bf Author Contributions}\\
 C.L., L.Z. and X.J.Z. proposed and designed the research. G.Y.D, X.C.W. and C.Q.J. contributed to LiFeAs crystal growth. C.L., Y.Q.C., Y.W., Q.G., G.D.L., Y.H., Q.Y.W., F.F.Z., S.J.Z., F.Y., Z.M.W., Q.J.P., Z.Y.X., L.Z. and X.J.Z. contributed to the development and maintenance of Laser-ARPES system. C.L. contributed to software development for data analysis. C.L., Y.Q.C. and Y.W. carried out the experiment. C.L., L.Z. and X.J.Z. analyzed the data. C.L., L.Z. and X.J.Z. wrote the paper. All authors participated in discussion and comment on the paper.

\newpage

\begin{figure*}[tbp]
\begin{center}
\includegraphics[width=1.0\columnwidth,angle=0]{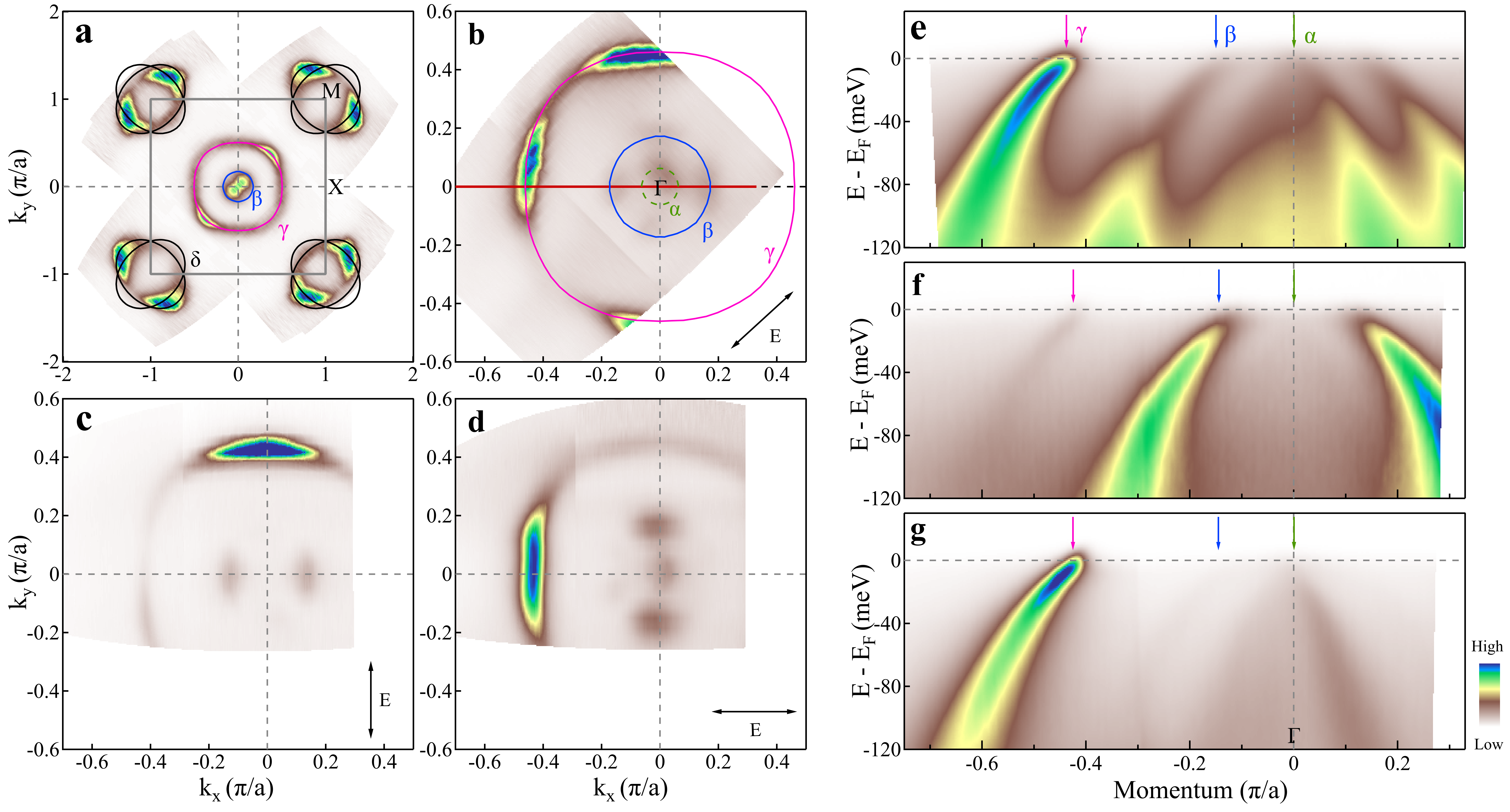}
\end{center}
\caption{\textbf{Electronic structure of LiFeAs measured under different polarization geometries.} (a) Overall Fermi surface of LiFeAs measured with a photon energy of 21.218 eV. Two Fermi surface sheets observed around $\Gamma$ are labeled as $\beta$ (blue circle) and $\gamma$ (pink circle) while two crossing elliptical Fermi surface sheets observed around M point are labeled as $\delta$ (black ellipses).  (b-d) Fermi surface of LiFeAs measured by using laser ARPES with a photon energy of 6.994 eV under different polarization geometries. The direction of the electric field vector E corresponding to the three polarization geometries are marked by double arrows in the bottom-right corner of each panel. We note that, while the electric field vector E in (c) fully lies in the sample plane, there is some component of the electric field vector E that is outside of the sample plane in (b) and (d).  In (b), the $\alpha$ band is also marked (dashed green circle) around $\Gamma$ in addition to the $\beta$ and $\gamma$ Fermi surface.  (e-g) The band structure of LiFeAs measured along the $\Gamma$-$X$ direction under three different polarization geometries that correspond to (b), (c) and (d), respectively. The location of the momentum cut is marked in (b) by a red line.  The green, blue and pink arrows point to the $\alpha$, $\beta$ and $\gamma$ bands, respectively.
}
\end{figure*}

\begin{figure*}[tbp]
\begin{center}
\includegraphics[width=1\columnwidth,angle=0]{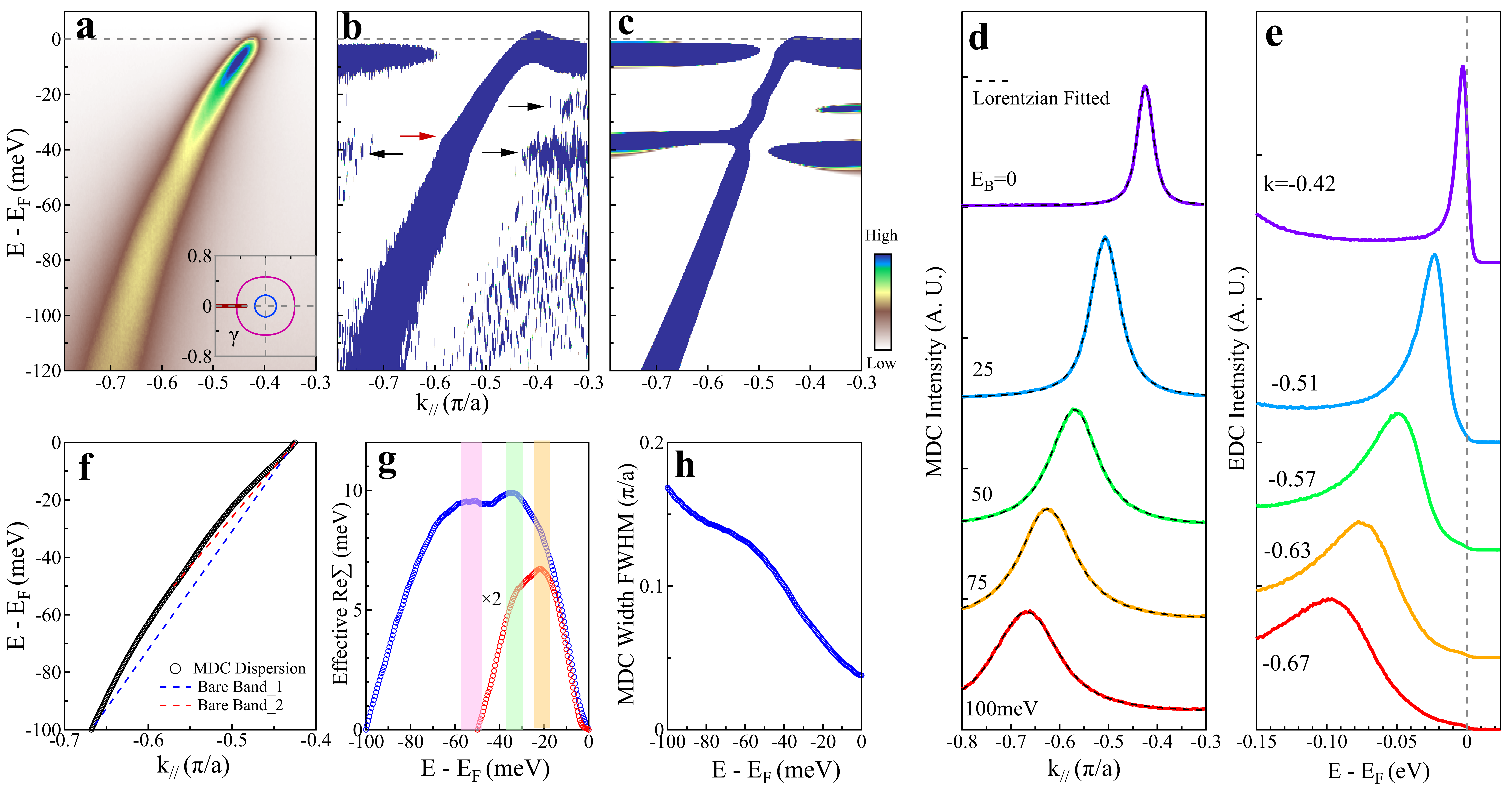}
\end{center}
\caption{\textbf{Electron dynamics of the $\gamma$ band of LiFeAs measured along the $\Gamma$ - $X$ direction at 20 K.} (a) The $\gamma$ band measured along the $\Gamma$ - $X$ direction.  The location of the momentum cut is marked by the red line in the inset.  (b) The second derivative image of (a) with respect to energy. (c) Second derivative image of the simulated single-particle spectral function which considers electron coupling with two bosonic modes at 20 meV and 34 meV.  (d) Momentum distribution curves (MDCs) at several representative binding energies. The MDCs are fitted by Loretzians that are overlaid as dashed lines on the measured data. (e) Representative energy distribution curves (EDCs) at several momenta. (f) Dispersion relation obtained by MDC fitting. The dashed red and blue lines represent empirical bare bands that are used to get the effective real parts of the electron self-energy $Re\Sigma$ (red line and blue line) shown in (g). The observed features are marked by pink, green and orange strips. (h) Corresponding MDC width (full width at half maximum, FWHM) of the $\gamma$ band in (a) from the MDC fitting.
}
\end{figure*}

\begin{figure*}[tbp]
\begin{center}
\includegraphics[width=1.0\columnwidth,angle=0]{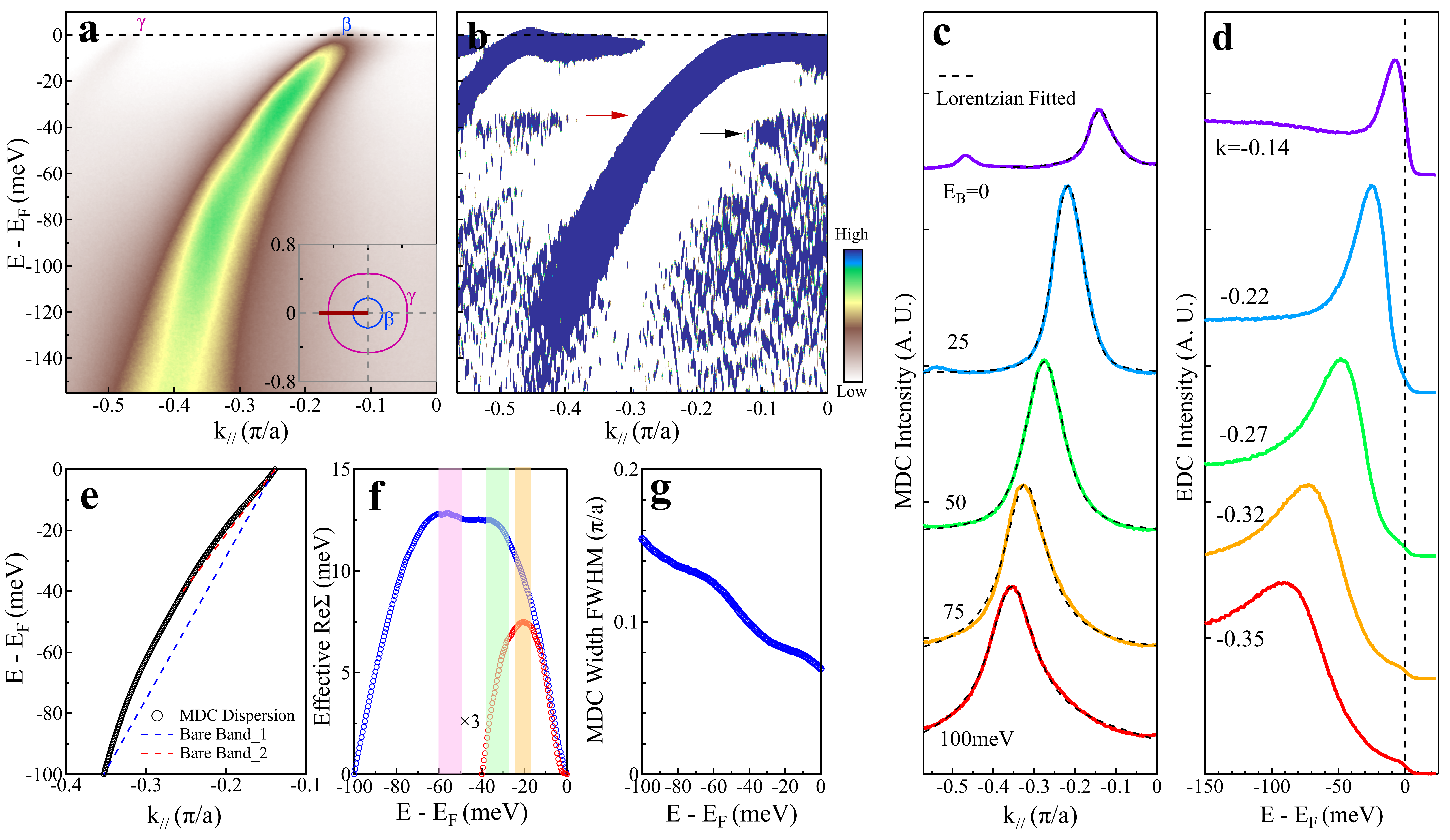}
\end{center}
\caption{\textbf{Electron dynamics of the $\beta$ band of LiFeAs measured along the $\Gamma$ - $X$ direction at 20 K.} (a) The $\beta$ band measured along the $\Gamma$ - $X$ direction.  The location of the momentum cut is marked by the red line in the inset.  (b) The second derivative image of (a) with respect to energy. (c) MDCs at several representative binding energies. The MDCs are fitted by Loretzians that are overlaid as dashed lines on the measured data. (d) Representative EDCs at several momenta. (e) Dispersion relation obtained by MDC fitting. The dashed red and blue lines represent empirical bare bands that are used to get the effective real parts of the electron self-energy $Re\Sigma$ (red line and blue line) shown in (f). The observed features are marked by pink, green abd orange strips. (g) Corresponding MDC width (FWHM) of the $\beta$ band in (a) from the MDC fitting.
}
\end{figure*}

\begin{figure*}[tbp]
\begin{center}
\includegraphics[width=1\columnwidth,angle=0]{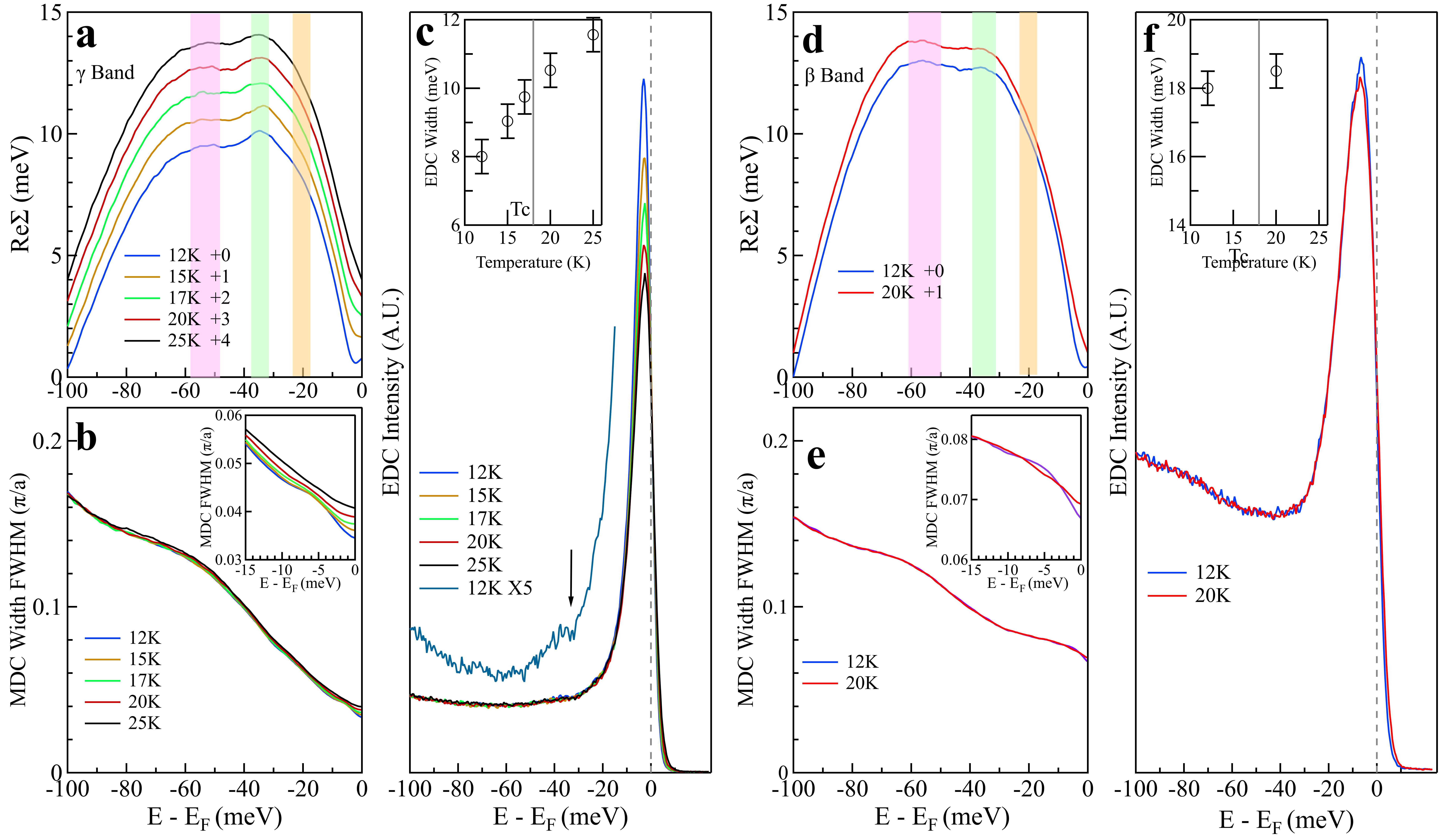}
\end{center}
\caption{\textbf{Temperature dependence of the electron dynamics for the $\beta$ and $\gamma$ bands in LiFeAs.} (a) Temperature dependent effective real part of electron self-energy of the $\gamma$ band.  For clarity, the curves are offset along the vertical axis.  (b) Corresponding MDC width of the $\gamma$ band measured at different temperatures. The upper-right inset shows the MDC width near the $E_{F}$ region.  (c) EDCs measured at the $k_{F}$ point of the $\gamma$ band at different temperatures.  The EDC at 12 K is also multiplied by 5 times to show the dip structure near 34 meV as marked by an arrow. The upper-left inset shows the temperature dependence of the EDC width (FWHM) of the $\gamma$ band.  (d) Temperature dependent effective real part of electron self-energy of the $\beta$ band.  The curves are offset along the vertical axis for clarity.  (e) Corresponding MDC width of the $\beta$ band measured at different temperatures. The upper-right inset shows the MDC width near the $E_{F}$ region.  (f) EDCs measured at the $k_{F}$ point of the $\beta$ band at different temperatures. The upper-left inset shows the temperature dependence of the EDC width (FWHM) of the $\beta$ band.
}
\end{figure*}


\end{document}